\documentclass[twocolumn,floats,floatfix,showpacs,amssymb,prd,superscriptaddress,nofootinbib]{revtex4-1}
\usepackage{graphicx, epsfig, amssymb}
\usepackage{epstopdf}
\usepackage{amsmath, amsfonts}
\usepackage{bm}
\usepackage[linktocpage]{hyperref}
\usepackage[caption=false]{subfig}
\usepackage[usenames]{color}

\begin{document}

\title{\large Gravitational Wave Memory from Gamma Ray Bursts' Jets}

\author{Ofek Birnholtz} \email{ofek.birnholtz@mail.huji.ac.il}
\affiliation{Racah Institute of Physics, Hebrew University, Jerusalem 91904, Israel.}

\author{Tsvi Piran} \email{tsvi@phys.huji.ac.il}
\affiliation{Racah Institute of Physics, Hebrew University, Jerusalem 91904, Israel.}

\date{\today}

\def\blue{}					

\newcommand{\pao}{\bf\color{red}}
\newcommand{\eb}[1]{\textcolor{blue}{[{\it\textbf{EB: #1}}]} }

\def\be{\begin{equation}}
\def\ee{\end{equation}}
\def\beq{\begin{eqnarray}}
\def\eeq{\end{eqnarray}}
\def\non{\nonumber}
\def\nn{\nonumber}
\def\L{\mathcal{L}}
\def\p{\partial}
\def\half{{\textstyle{1\over2}}}
\def\ss{\scriptscriptstyle}
\renewcommand{\vec}[1]{\boldsymbol{#1}}
\newcommand{\us}[1]{\textcolor{green}{[{\it\textbf{US: #1}}]} }
\newcommand{\vc}[1]{\textcolor{blue}{[{\it\textbf{VC: #1}}]} }

\def\IInt{\int\!\!\!\!\!\int}
\def\IIInt{\int\!\!\!\!\!\int\!\!\!\!\!\int}

\def\a{\alpha}
\def\bt{\beta}
\def\d{\delta}
\def\D{\Delta}
\def\eps{\epsilon}
\def\veps{\varepsilon}
\def\Gm{\Gamma}
\def\gm{\gamma}
\def\k{\kappa}
\def\l{\lambda}
\def\L{\Lambda}
\def\o{\omega}
\def\O{\Omega}
\def\vph{\varphi}
\def\vpi{\varpi}
\def\s{\sigma}
\def\S{\Sigma}
\def\vsgm{\varsigma}
\def\t{\theta}
\def\T{\Theta}
\def\vt{\vartheta}
\def\z{\zeta}
\def\X{\times}


\newcommand{\refeq}[1]{eq.~(\ref{#1})}
\newcommand{\refig}[1]{fig.~(\ref{#1})}
\newcommand{\Refeq}[1]{Eq.~(\ref{#1})}
\newcommand{\Refig}[1]{Fig.~(\ref{#1})}

\newcommand{\sect}[1]{\setcounter{equation}{0}\section{#1}}
\renewcommand{\theequation}{\thesection.\arabic{equation}}

\newcommand{\tab}{\hspace*{3em}}

\setcounter{section}{0}
\setcounter{equation}{0}


\begin{abstract}
While the possible roles of GRBs' progenitors as Gravitational Waves (GW) sources  have been extensively studied, little attention has been given to the GRB jet itself as a GW source. We expect the acceleration of the jet to produce a Gravitational Wave Memory signal. While all relativistic jet models display anti-beaming of GW radiation away from the jet axis, thus radiating away from directions of GRBs' $\gm$ radiation, this effect is not overwhelming. The decrease of the signal amplitude towards the cone of $\gm$-ray detectability is weak, and for some models and parameters the GW signal reaches a significant amplitude for much of the $\gm$-ray cone. Thus both signals may be jointly detected. We find different waveforms and Fourier signatures for uniform jets and structured jet models - thus offering a method of using GW signatures to probe the internal structure and acceleration of GRB jets. The GW signal peaks just outside the jet (core) of a uniform (structuted) jet. Within the jet (core) the GW signal displays wiggles, due to a polarization effect; thus for a uniform jet, the peak amplitude accompanies a smoother signal than the peak of a structured jet. For the most probable detection angle and for typical GRB parameters, we expect frequencies $\lesssim600Hz$ and amplitudes $h\sim10^{-25}$. Our estimates of the expected signals suggest that the signals are not strong enough for a single cluster of DECIGO nor for aLIGO's sensitivities. However, {\blue sensitivies of $\sim10^{-25}/\sqrt{Hz}$ in the DECIGO band should suffice to detect typical long GRBs at $2Gpc$ and short GRBs at $200Mpc$, implying a monthly event of  a long GRB and a detection of a short GRB every decade. In addition, we expect much more frequent detection of GW from GRBs directed away from us, including orphan afterglows. The ultimate DECIGO sensitivy should increase the range and enable detecting these signals in all models even to high cosmological z}.
\end{abstract}

\maketitle


\sect{Introduction}							%

General Relativity predicts Gravitaional Waves (\cite{Einstein}), and evidence of their effect on astrophysical systems has established their existance (\cite{HulseTaylor,Hulse,Taylor}). However, no direct observation of such waves has yet been achieved. The searches for gravitational wave sources (\cite{Weber}) can be simplified if one knows in advance and with enough precision where and when to look. When searching for a GW signal amidst background noise, it is also helpful to know external parameters regarding the source that might help characterize it and its waveform-template. In addition, the ability to study a single astrophysical event simultaneously both in the EM spectrum and in its GW spectrum can provide much more information and insights than the mere sum of the separate measurements. For these reasons, it is interesting to look for Astrophysical objects that emit both EM and GW radiation, and it is interesting to explore whether and under what conditions they are jointly detectable (\cite{Eichler,Kochanek,Fireball,Piran,Kocsis06,Kocsis08,Lang,Bloom,Demorest,Jenet,Phinney,Schnittman}). GRBs are the most natural candidates observed to emit strong EM radiation and expected to emit GW radiation as well \cite{Kochanek}. In particular, short GRBs are expected to be associated with merging binary neutron stars, which are the classical candidates for detection by advanced GW detectors (\cite{Paczynski,AbbotGRBGW1,AbbotGRBGW2,Hough,Corsi}).
\\
Different proposed mechanisms concerning GRB's have been suggested and studied as possible sources for Gravitational Waves. However, little attention has been given to the relativistic jet itself as a source. Regardless of the progenitor type, all GRB's share an initial phase of an accelerating relativistic jet. GRB's involve explosions which release about $E_j\sim10^{51}$ergs of energy in accelerating jets (a forward jet and a backward jet), reaching ultra-relativistic velocities with Lorentz factors $\Gm\geq$100 , or even up to 1000 \cite{Abdo}. The jets are strongly beamed into narrow (double) cones, of angle $\t_{0}\sim0.1$. The duration of the observed bursts ranges from less than 0.01sec to more than 100sec, while their light curves show rapid variability, at times on scales of about 0.1sec. The ultra-relativistic non spherically-symmetrical acceleration of this energetic jet is expected to produce Gravitational Radiation (\cite{Maggiore,Segalis,Sago,Hiramatsu,Favata,Rezzolla,Grishchuk,Madau,Miller,Prince}). In this work, we focus on this GW signal, from a single pulse of an accelerating jet, regardless of the progenitor.\\
We model the jet in 3 stages of increasing complexity: first, as an instantaneous infinitely narrow jet pulse (the ``point particle approximation", section (2)) that enables us to estimate the signal strength; second, as an instantaneous cone with a finite width and mass distribution (aggregate models, section (3)) for an understanding of the angular distribution; and finally as a cone of both finite width and finite time (a prolonged acceletation model, section (4)) for am estimation of the temporal structure (waveforms). We analyze the results of each model, determining the directions of radiation (and in particular the possibility of a joint detection of GW and EM ($\gm$-ray)), and characteristic signal amplitudes, energy output, and frequency behaviour. We build each model using results from the simpler ones. In section (5) we discuss the possibility of detection, and future work.\\
{\bf Note on conventions}: Adhering the General Relativists' convention of ``natural units", we use (c=G=1). However, certain formulae for comparison with physical results and scales employ the cgs system. The space time metric signature is $diag(-+++)$.

\sect {Instant acceleration of a point-particle}				%

\subsection{GW Memory and the Point Particles Approximation} 
While the progenitors emit quasi-periodical gravitational waves (Coalescing of Compact Binaries, etc.), the acceleration of the jet itself is prompt and non-periodic, and thus we do not expect ``waves". Rather, we expect the jet to produce a gravitational wave memory (GWM; or Zero Frequency Limit, ZFL \cite{Segalis,Sago,Hiramatsu,Favata,Rezzolla,Grishchuk,Madau,Miller,Prince,Smarr}), defined as:
\begin{equation} \label{hlim}
\D h^{mem}_{+,\X}=\lim_{t \to +\infty} h_{+,\X}(t) - \lim_{t \to -\infty} h_{+,\X}(t),
\end{equation}
where $h_+$ and $h_\X$ denote the plus and cross standard polarizations of the wave function \cite{Misner}.\\
Einstein's equation reduces - in the regime of linearized gravity and under the harmonic gauge condition \cite{Weinberg} - to a wave eqaution for the metric perturbation ($h_{\mu\nu}=g_{\mu\nu}-\eta_{\mu\nu}$), with the Stress-Energy tensor serving as the source term: 
\begin{equation} \label{linearizedgravity}
\Box h_{\mu\nu}=-16\pi S_{\mu\nu}=-16\pi (T_{\mu\nu}-\frac{1}{2}\eta_{\mu\nu}T^\lambda_\lambda).
\end{equation}
\\
The solution of this equation is:
\begin{equation} \label{linearizedgravitysolution}
h_{\mu\nu}({\bf x},t)=4\int\!\! d^3{\bf x'} \frac{S_{\mu\nu}({\bf x'},t-|{\bf x}-{\bf x'}|)}{|{\bf x}-{\bf x'}|}.
\end{equation}
The simplest approximation of a Stress-Energy tensor for a GWM calculation is obtained using the ``Collision Approximation", which models the process as an instantaneous exchange of ingoing point-particles with outgoing point-particles. The gravitational signal is derived from the incoming/outgoing momenta of the $n$ particles ($P_n^\mu$/$P_n^{'\mu}$) giving the stress-energy tensor $T^{\mu\nu}$:
\begin{equation} \label{stressenergytensor}
\begin{array}{lll}
T^{\mu\nu}(x,t)&=&
\sum_n\frac{P_n^\mu P_n^\nu}{E_n}\d^3({\bf x}-{\bf v_n}t)\t(-t)\\
&&+\sum_n\frac{P_n^{'\mu} P_n^{'\nu}}{E'_n}\d^3({\bf x}-{\bf v'_n}t)\t(+t).\\
\end{array}
\end{equation}
Combining these two equations and integrating over the $\delta$-functions gives the gravitational signal. Similar ``Collision Approximation" methods have been used by Weinberg\cite{Weinberg}, Piran\cite{Piran} and others (\cite{Sago}-\cite{Rezzolla},\cite{Braginsky},\cite{Thorne}). Smarr \cite{Smarr} has shown that this form reproduces (using momentum conservation) the gravitational wave signal form used by Braginsky, Thorne and others for instantaneous collisions (of $N$ particles with rest-masses $m_A$ and velocities $v_A$ at angles $\t_A$ and distance $r$ to the observer, and with the $\D$ marking the memory effect as defined in \refeq{hlim}):
\begin{equation} \label{hpointparticle1}
\begin{array}{lll}
\D h^{TT}_{jk}&=&
\sum_{A=1}^N \frac{4m_A}{r\sqrt{1-v_A^2}}\left. \left[\frac{v_A^j v_A^k}{1-v_A\cos\t_A} \right]^{TT}\right|^{outgoing}_{incoming} \\
&=&
\sum_{A=1}^N \frac{4\Gm_A m_A}{r}\left. \left[\frac{\beta_A^j \beta_A^k}{1-\beta_A\cos\t_A} \right] ^{TT}\right|^{outgoing}_{incoming}\\
\end{array}
\end{equation}
This is a purely quadrupole radiation pattern, analogously to the Lienard-Wiechert Potentials for electromagnetic radiation \cite{Jackson}.

\subsection{The Jet as a Point-Particle}		
\begin{figure}[h]
\includegraphics[width=60mm]{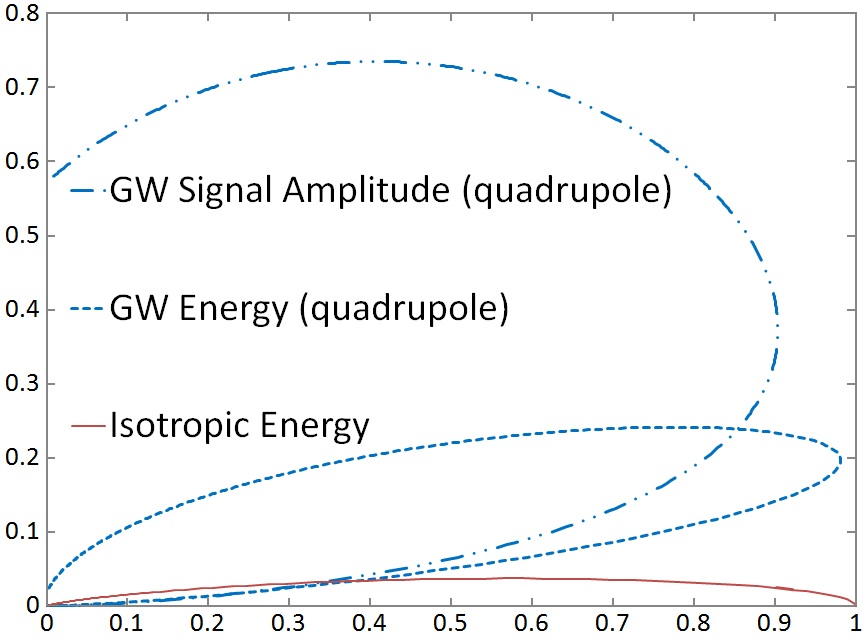}
\caption{Antenna Diagram showing relative beamed directional intensity (energy) for isotropic (red) and quadrupole (blue, dotted) radiation from a point source (infinitesimal width), for $\Gm=7, \beta=0.99$. The quadrupole radiation pattern is the expected for GW radiation from a point source.  The diagram also shows the relative directional signal amplitude of the GW signal (blue, dash-dotted).}
\label{pointparticle}
\end{figure}

The simplest approximation of the jet's GW signal would be to treat the entire jet as a single point-mass undergoing an instantaneous linear acceleration from rest to $\Gm$, reaching the total jet energy $E_j$. This model is of course over-simplified, as it neglects both the jet's finite width (the model assumes it to be infinitely concentrated and narrow) and the finite duration of the acceleration phase (which the model assumes to be instantaneous). However, it provides an understanding of the relevant characteristic signal and output energy scales (computed from the accelerated jet energy $E_j$), as well as an approximate angular distribution.\\
The change induced in the metric between $t\to\pm\infty$ for an observer at a distance r away is determined by the angles between his line-of-sight and the various velocities, the distance and the energy of the jet. As in \cite{Sago,Hiramatsu} we set N=2 for an initial ``incoming particle" at rest ($\beta_i=0$) and a final ``outgoing particle" of mass $m_f=m=E_j/\Gm$ with $\beta_f=\beta(\sin\t\cos\phi, \sin\t\sin\phi, \cos\t)$, where $\bt$ and the angles are in the observer frame (the $z$ axis points to the observer). The contribution of a point-particle accelerating instantaneously from 0 to $\beta$ to the gravitational perturbation $h$ is
\begin{equation} \label{hpointparticledeltah}
\D h^{pp}=\D h_+ + i\D h_\X=\frac{2\Gm m\beta^2}{r}\frac{\sin^2\t}{1-\beta\cos\t}e^{2i\phi}.
\end{equation}
We note here that for relativistic velocities ($\beta\sim1$) the typical scale for the amplitude of the wave will be determined by
\begin{equation} \label{htypicalamp}
h_{typ}\sim(\frac{2\Gm M}{r}).
\end{equation}
We see that the angular distribution of radiation from an instantaneously accelerating point source is set by two competing effects. {\blue The transformation from the CMRF to the observer frame Lorentz-contracts the radiation forward, by a factor of $(1-\beta\cos\t)^{-1}$. This beaming (as it would affect isotropic emission) is shown in \refig{pointparticle}. However, the multipole pattern of the radiation, expressed as $\sin^n\t$ (for multipole order $n$), directs the radiation away from the forward direction.} For the GW quadrupole radiation ($h_{GW}\propto\sin^2\t/(1-\beta\cos\t)$) the result is ``anti-beaming" away from the forward direction (\refig{pointparticle}), i.e. an almost uniform distribution outside a forward beam of width $\t\sim\sqrt{2}\Gm^{-1/2}$ (for which $1-\beta cos\t \sim\Gm^{-1})$ \cite{Maggiore}. This implies detection of both EM and GW radiation requires attributing a finite width to the jet.

\subsubsection{The energy flux}			
The energy carried by the GW depends on the perturbation amplitude as well as on the frequency. We have calculated the GW Memory (ZFL) effect of a GRB jet, considering only initial (rest) and final (accelerated) particle momenta, or equivalently, assuming instantaneous acceleration. Thus for energy calculations we attribute a finite timescale for the acceleration $\d T$ in the source frame. In the observer frame, 
\begin{equation} \label{radiatedpower}
\frac{dE_{GW}}{d\omega' d\Omega}=\frac{r^2}{16\pi^2}(\D h)^2,
\end{equation}
and by integrating over observed frequencies up to $\frac{2\pi}{\d T'(\t)}$, we find the radiated GW energy
\begin{equation} \label{radiatedenergy}
\frac{dE_{GW}}{d\Omega}=\frac{r^2}{8\pi}\frac{(\D h)^2}{\d T'(\t)},
\end{equation}
where the observed timescales are related both to the acceleration timescales and to the observation angle by
\begin{equation} \label{timecontraction}
\d T'(\t)= \d T(1-\beta cos\t),
\end{equation}
as pointed out by Segalis \& Ori \cite{Segalis}. Thus the angular dependence of the energy flux is (suppressing the prefactor $\frac{2\Gm m\beta^2}{r}$ and factors of G and c)
\begin{equation} \label{Epointparticledeltah2}
\frac{dE_{GW}}{d\Omega}\propto \frac{|\D h^{pp}|^2}{\d T'}\propto\frac{\sin^4\t}{(1-\beta\cos\t)^3}.
\end{equation}
The additional $(1-\beta\cos\t)$ term in the denominator enhances the beaming of the energy flux in the forward direction, and its peak is still outside a $\t\sim\sqrt{2}\Gm^{-1}$ forward cone (\refig{pointparticle}).

\subsubsection{Physical Parameters}		
Assuming a physical time-scale for the acceleration process, $\d T\sim30msec$\footnote {we assume a stage of outward baryonic expansion from $R\sim10^7cm$ to $R\sim10^9cm$, where for most of the expansion the velocity $v\sim c$.}, we expect a waveform displaying typical frequencies $f\sim30Hz$ in angles away from the axis of acceleration. As the angle is closer to the acceleration axis, the typical frequencies are larger, inversely to the time contraction (\refeq{timecontraction}). This places these sources in the frequency band for detection with the earth-based detectors (such as LIGO, Virgo and for smaller frequencies - DECIGO and  NGO).\\
We consider characteristic jets reaching total kinetic energy $E^j\sim10^{51}ergs$ and a final Lorenz factor $\Gm\sim100$. This sets the normalization condition for the mass parameter from (\refeq{hpointparticledeltah}, (\ref{htypicalamp})). Integrating over all solid angles and restoring the physical units and parameters (\refeq{htypicalamp}), the total output is:
\begin{equation} \label{totaloutput}
\begin{array}{lll}
E_{GW}=&\frac{G}{c\d T}M^2\int d\cos\t\frac{\D h^{jet}(\t))^2}{1-\beta cos\t}.
\end{array}
\end{equation}
Using $O(1)$ estimates for $\D{}h^{jet}(\t_v)$ outside the narrow forward cone, and substituting the physical parameters (for $r\sim500Mpc)$, we estimate the energy and signal amplitude as:
\begin{equation} \label{EGW_pp}
E^{PP}_{GW}\sim 10^{44}erg \biggl(\frac{M}{10^{51}erg}\biggr)^2 \biggl(\frac{\d T}{30ms}\biggr)^{-1},
\end{equation}
\begin{equation} \label{hpeak_pp}
h^{PP}_{peak}\sim 10^{-25} \biggl(\frac{M}{10^{51}erg}\biggr) \biggl(\frac{r}{500Mpc}\biggr)^{-1}.
\end{equation}

\sect{Instantaneous acceleration of an Aggregate Source}		%

\subsection{Aggregate Source Models}				
\begin{figure}[h]
\includegraphics[scale=0.40]{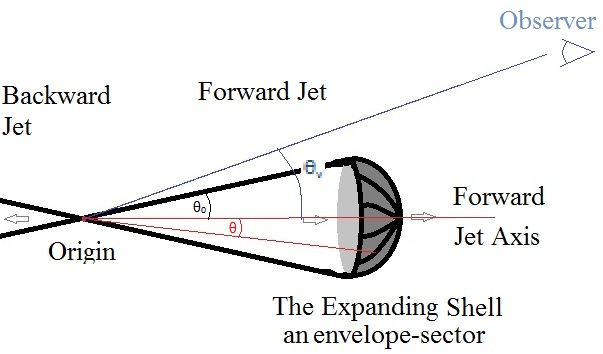}
\caption{Our Jet modelled as a sector of an envelope of mass-energy, expanding.}
\label{jetdiagram}
\end{figure}

We turn now to study a slightly more realistic model: an aggregate model of an expanding shell (a sector of an envelope, \refig{jetdiagram}). The ejected mass is now distributed in a shell according to some angular distribution $f(\t)$ that is axis-symmetrically around the jet's central axis. The jet axis is at an angle $\t_v$ relative to the observer's line-of-sight (the observer is at a distance $r$ away). We treat every mass-element in the shell as an accelerated point-particle, and find its contribution $\D{}h^{pp}$ to the signal received at $r$. For every $\t_v$ we integrate the contributions $\D{}h^{pp}$ over ($\t,\phi$), to find the total signal from the jet, $h^j$:
\begin{equation} \label{deltahjet}
\D{}h^j(\t_v)=\int f(\t,\phi)\D{}h^{pp}(\t,\phi,\t_v) \sin\t d\t d\phi,
\end{equation}
where $f$ is a normalized for conservation of total ejected energy between the different models and parameters. We consider two possible models for the mass distribution in the jet (Both models are of course idealized approximations to the real, unknown angular distribution of a GRB jet):\\
a. A uniform jet:
\begin{displaymath}
f(\t) = \left\{
\begin{array}{ll}
   1 & \t < \t_0,\\
   0 & \t > \t_0.\\
\end{array}
\right.
\end{displaymath}
\\
b. A structured jet (following \cite{Rossi}):
\begin{displaymath}
f(\t) = \left\{
\begin{array}{ll}
   1				& \t <\t_{core}=\Gm^{-1},\\
   (\Gm\t)^{-2}	& \t_{core} < \t < \t_0,\\
   0				& \t_0 < \t.\\
\end{array}
\right.
\end{displaymath}

\subsection{Results}				
\begin{figure}[h]
\includegraphics[width=86mm]{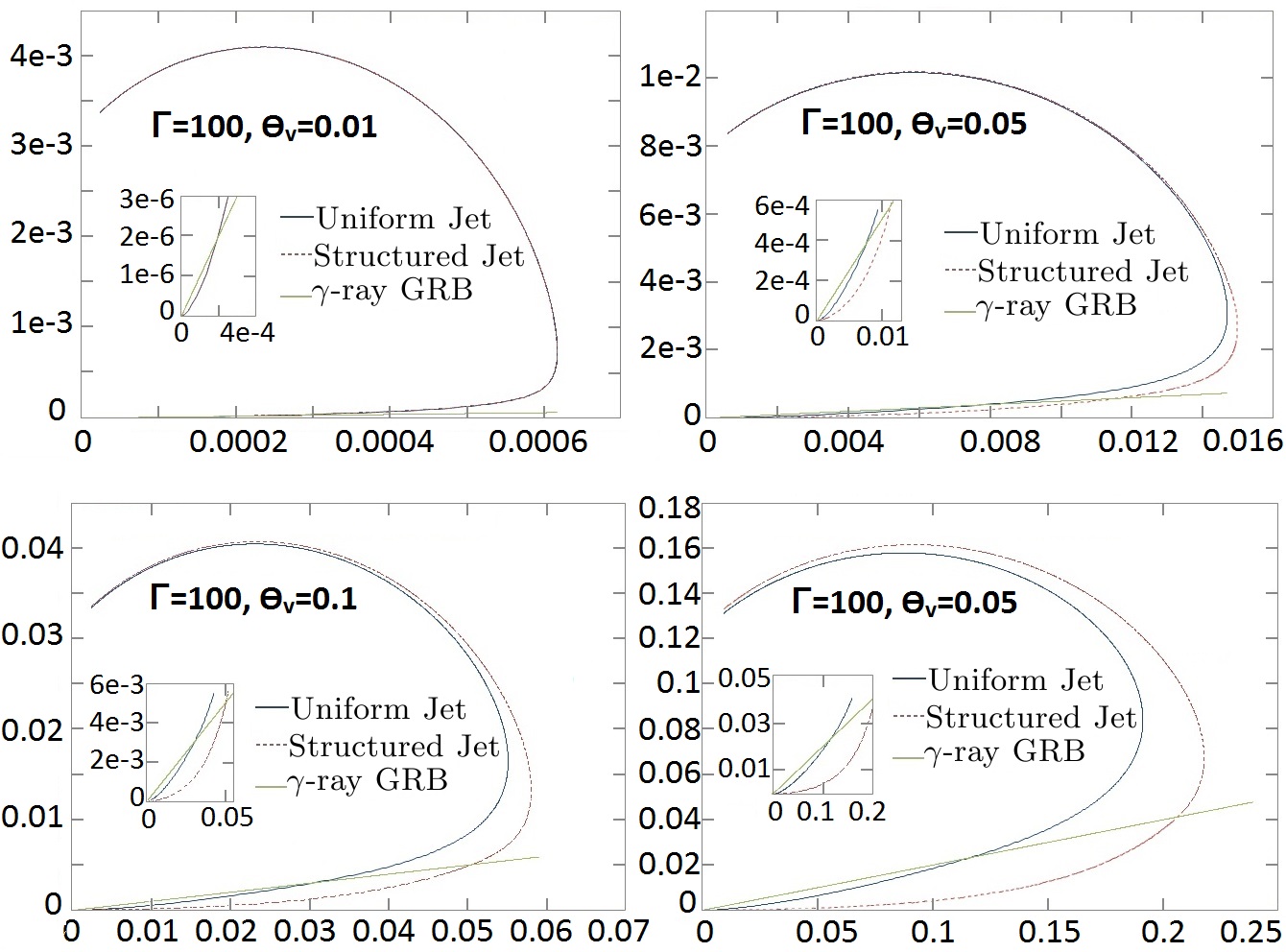}
\caption{Antenna diagrams showing angular distribution of the GW signal amplitude emitted from $\Gm=100$ for uniform and structured jets of different jet opening angles $\t_0 (0.01,0.05,0.1,0.2)$. The axes mark relative directional amplitude, the x-axis marks the jet axis and the y axis is perpendicular to it. The blue curve describes the uniform jet, while the red describes the structured jet. Each pair is normalized to the same total jet energy, demonstrating the relative wave intensities and angular distributions. Both models coincide for $\t_0=0.01$; for larger $\t_0$, the structured jets are narrower and their radiation is both stronger and it is beamed into a sharper emission cone. The green line demarcates the opening angle of the jet itself, approximating the region of detectability of the $\gm$-ray signal from the GRB jet. Small angles are zoomed-in (boxed), showing the regions of joint detectability.}
\label{jetantena}
\end{figure}

\begin{figure}[!hbp]
\includegraphics[width=86mm]{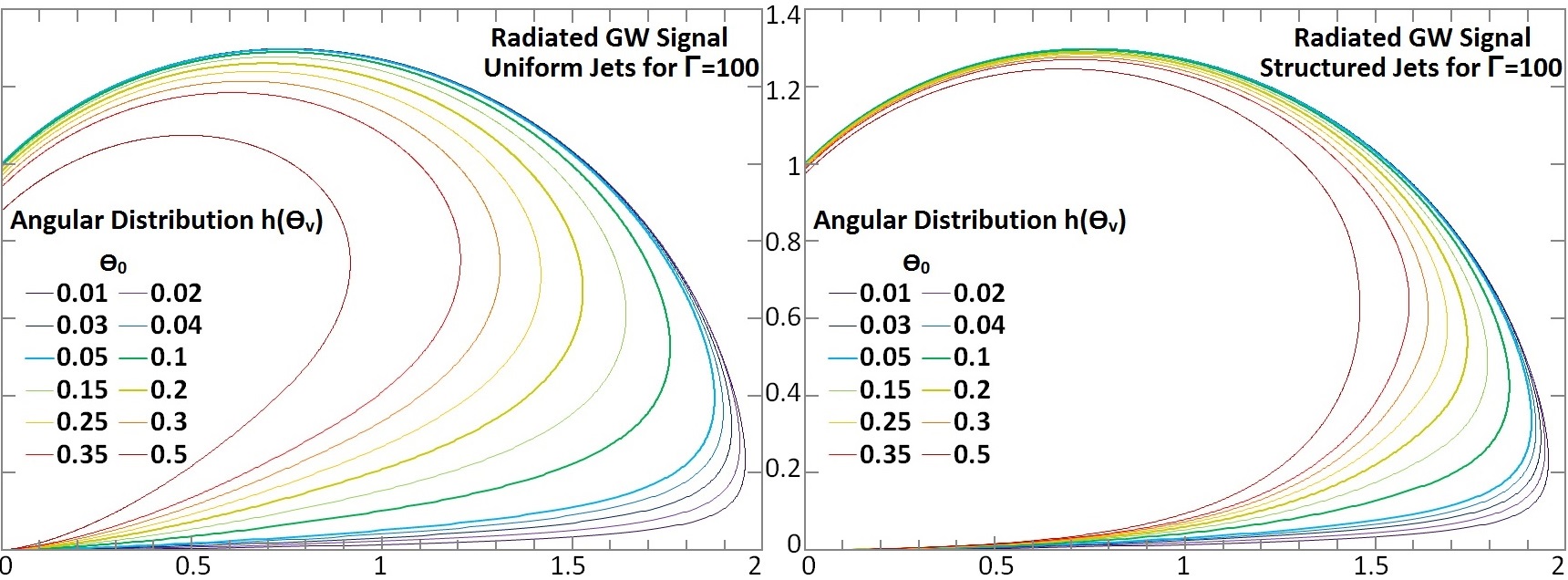}
\caption{Antenna diagrams for the signal amplitude of GW emitted from uniform (left) and structured (right) jets of different opening angles $\t_0$. All are normalized to the same total jet energy $10^{51}ergs$. The axes mark relative directional amplitude, the x-axis marks the jet axis and the y axis is perpendicular to it. The radiation pattern becomes wider, weaker, and more anti-beamned as $\t_0$ increases. Both models coincide for $\t_0=0.01$.}
\label{both}
\end{figure}

We have numerically calculated the integral \refeq{deltahjet} for both jet models with the same total jet kinetic energy of $E_j=10^{51}ergs$. \Refig{jetantena} depicts a graphical comparison betwenn the angular distribution of the GW signal for uniform and structured jets, for typical parameters ($\Gm=100; \t_0=0.01, 0.05, 0.1, 0.2$). Fig. (\ref{both}a) and (\ref{both}b) depict together the energy and angular distributions for uniform and structured jets of different $\t_0$ (respectively). Note that the minimal openning angle considered $0.01$ corresponds to the narrowest possible jet, for which $\t_0 =\Gamma^{-1}$.

\begin{figure}[h]
\includegraphics[width=86mm]{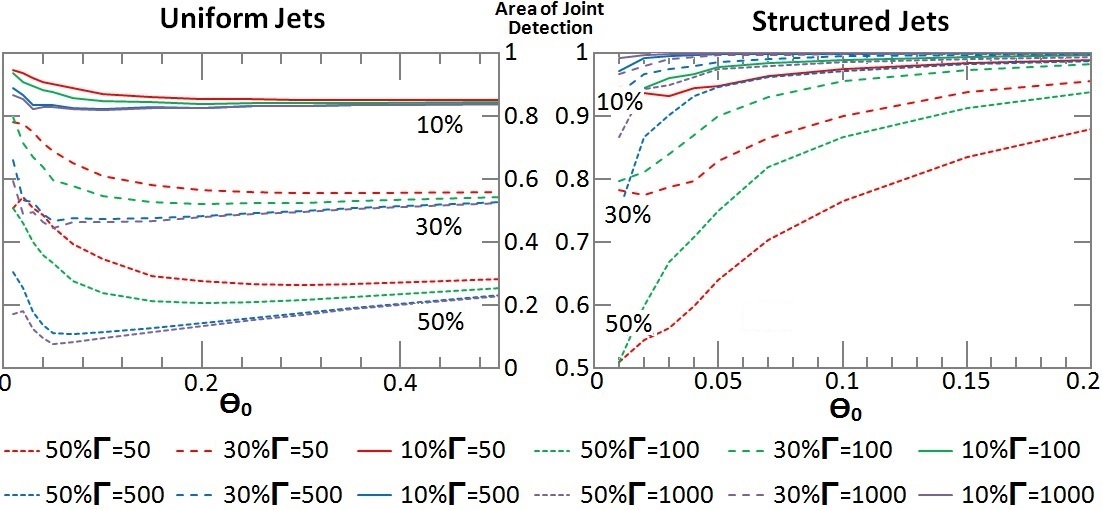}
\caption{The region of a possible joint detection of a GW signal with $\gm$-ray signal, relative to the total $\gm$-ray detection cone. The left figure describes uniform jets, structured jets are on the right. For each $\Gm$ value the graph shows the results for 3 values of the threshold signal amplitude (rel. to the peak amplitude): 50\% (bottom, dotted), 30\% (middle, dashed) and 10\% (top, continuous). Different $\Gm$ are color-coded as red ($\Gm=50$), green ($\Gm=100$), blue ($\Gm=500$) and purple ($\Gm=1000$)}
\label{jointdetection}
\end{figure}

Combining our numerical results for $\D{}h^{jet}(\t_v)$ with all of the physical parameters (for $r\sim500Mpc$, using \refeq{htypicalamp}), we find the jet width's effects on the emitted signal (compare with (\ref{hpeak_pp})). We find that for a uniform jet:
\begin{equation} \label{hpeak_uni}
h^{UNI}_{peak}\sim 2\cdot10^{-25} \biggl(\frac{M}{10^{51}erg}\biggr) \biggl(\frac{\t_0}{0.01}\biggr)^{-0.1} \biggl(\frac{r}{500Mpc}\biggr)^{-1},
\end{equation}
while for a structured jet:
\begin{equation} \label{hpeak_struc}
h^{STRUC}_{peak}\sim 2\cdot10^{-25} \biggl(\frac{M}{10^{51}erg}\biggr) \biggl(\frac{\t_0}{0.01}\biggr)^{-0.04} \biggl(\frac{r}{500Mpc}\biggr)^{-1}.
\end{equation}
For the common parameter of $\Gm=100$, our models for a uniform and a structured jet coincide for $\t_0=\t_{core}=\Gm^{-1}=0.01$. However, the signal strength and energy decrease as $\t_0$ increases, with the uniform wave decreasing faster. This result is also evident in \refig{jetantena},(\ref{both}). This result is general: for any $\t_0>\t_{core}=\Gm^{-1}$, we expect the wave pattern from a structured jet to be stronger than the one from a uniform jet with the same paramers, and we expect the GW from uniform jets to decrease faster with $\t_0$ than the corresponding structured ones. This is due to the higher spherical-symmetric character of the uniform jet: for $\t_0>\Gm^{-1}$, the structured jet is always more focused than its uniform counterpart, and as $\t_0$ increases, it remains more focused than the uniform.\\
We defer more thorough study of possible detection by currently planned GW detectors (advanced LIGO, NGO, DECIGO, etc.), but do note that these signal amplitudes are weak (advanced LIGO \cite{LIGO} and probable modes of operation of DECIGO \cite{DECIGO,DECIGO2} offer strain sensitivities of $\sim10^{-24}-10^{-23}$). Thus, the exact parameters of the jet, as well as the jet's internal structure model, might affect the possibility of detection.

\subsection{The Angular Distribution and Possible Joint Detection}	
The GW signal from a single point-source is anti-beamed away from the axis of acceleration (at angles $\t_{peak}\gtrsim\Gm^{-1/2}$). The relation between the detection cone for the GRB's $\gm$-ray signal and the GW detection range for the uniform and structured aggeragte models is more complicated (fig. (\ref{jetantena},\ref{both}a,\ref{both}b)). For a very narrow uniform jet ($\t_0\lesssim\Gm^{-1}$), the angular spread is similar to that of a point source, and $\t_{peak}\gtrsim\Gm^{-1/2}$. The peak angle increases (away from the jet axis) as $\t_0$ increases, and for wider unifrom jets we find approximately $\t_{peak}^{UNI}\propto\t_0^{0.46}$. Thus for all $\t_0$ the uniform jet GW signal peaks above both $\t_{peak}\gtrsim\Gm^{-1/2}$ and $\t_0^{-1/2}$. This is a consequence of anti-beaming of the GW radiation away from the source axis. As the $\gm$-rays are only visible up to an observer angle of $\t_0+\Gm^{-1}$, which is smaller than $\t_{peak}$, the peak of the GW radiation is generally outside of the $\gm$-ray detection cone. The beaming effect is more pronounced at higher values of $\Gm$; but it is qualitatively the same. Thus the GW peak cannot be observed by an observer that detects the $\gm$-ray signal.\\
However, the fall-off of the GW signal towards the jet axis (and the cone where its $\gm$-rays are visible) is slow, as each of the point-particles' quadrupole signal itself is broad. Therefore if we are within the jet's $\gm$-ray cone, we might still see the corresponding GW signal at less-than-peak amplitude. For a uniform jet detecting the peak $\gm$-ray signal is possible anywhere within the $\gm$-ray cone (up to maximum angle $\t_0+\Gm^{-1}$ from the axis). The range of possible joint detectability, given as relative area (area of joint detectability region / total area of $\gm$-ray cone) is shown in \refig{jointdetection}.\\
The situation is very different for structured jets. While we still consider their $\gm$-ray signal to be as detectable anywhere up to $\t_0+\Gm^{-1}$, the GW signal comes mostly from the $\Gm^{-1}$ core, thus its peak is closer to the jet axis ($\t_{peak}^{STRUC}\propto\t_0^{0.35}$) and the signal fall-off away from the peak is slower. While in both models the radiation pattern becomes wider, weaker, and more anti-beamned as $\t_0$ increases, we find that this change is less pronounced for structured jets (\refig{both}). As \refig{jointdetection} shows, this implies a higher possibility of detection for structured jets, for the same parameter space.\\
Having examined ranges of $\Gm$ values (10..1000), $\t_0$ values (0.01..0.5), and relative threshold amplitudes (10\%, 30\%, 50\%) we find that even for uniform jets the result is quite optimistic, allowing joint detection for much of the $\gm$-ray cone area (approximately 80-90\%, 50-80\%, 20-50\% corresponding to the 10\%,30\%,50\% thresholds). Naturally the lower the threshold, the greater is the probability for joint detection. But even for a high 50\% threshold (for which we loose only a factor of 2 in the detection rate) we still expect a considerable joint detection probability. We note also the dependence of the joint detection probability on $\Gm$ and on $\t_0$: for smaller relativistic $\Gm$ factors, beaming/anti-beaming effects are less pronounced, and the possibility of a joint detection is larger. Regarding the opening angle $\t_0$, we find the largest probabilities for a joint detection for very small values of $\t_0$ (approaching point sources; and also implying stronger GRBs). The probability decreases as $\t_0$ increases, up to an angle $\sim \Gm^{-1/2}$, and then it rises very slowly rises. For structured jets, the parameter range of joint detection is even wider, allowing detection even of signals at the 50\% threshold at over 80\% of the $\gm$-ray cone for reasonable models ($\t_0\gtrsim0.1$), and over a larger fraction of the range for larger $\t_0$, as shown in \refig{jointdetection}.\\
For both models, the availability of an EM trigger ($\gm$-ray) setting the time and position in the sky of the source can significantly increases the sensitivity of the GW detectpr as compared to a random search \cite{Kochanek}, thus increasing the chances of a joint detection.


\sect {A prolonged Acceleration Model}					%
\subsection{The Model}							
\begin{figure}[h]
\includegraphics[width=86mm]{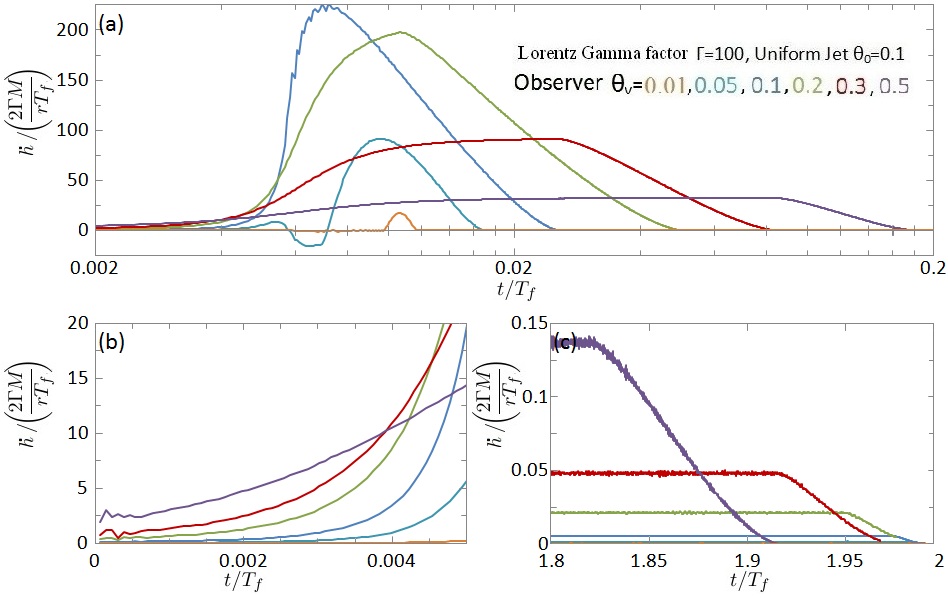}
\caption{$\dot{h}$ as a function of time for different observeration angles $\t_v$;$\Gm=100$, $\t_0=0.1$. (a) The waveform on a logarithmic time scale. The peaks are visible around $0.5\Gm^{-1}T_f$. (b) A close-up of the early signal, which is visible sooner the further off from the line of acceleration. (c) The late signal ($t>1.8T_f$) continues to rise very moderately, due to the backwards jet. The final drop in $\dot{h}$ marks to the termination of the jet expansion, at $2T_f$.}
\label{waveformsdeltah}
\end{figure}

So far we have considered jet models of different angular distributions, undergoing instantaneous acceleration from rest ($\beta=0,\Gm=1$) to $\beta\sim c,\Gm\sim100$ (following \cite{Piran,Segalis,Sago,Braginsky,Thorne}). We used the duration of the acceleration $\delta T$ just to estimate the typical frequencies. However, the instantaneous model is of course nonphysical, and in order to calculate waveforms, as well as the GW spectrum, we must examine more specific models for the acceleration, where higher multipoles than the quadrupole appear.\\
Our treatment of the GW radiation is analogous to the treatment of EM radiation in terms of Lienard-Wiechert Potentials (as in \cite{Jackson}), where the radiation contribution of an instantaneously accelerating point-particle source (with all quantities evaluated at retarded time) is given by
\begin{equation} \label{lienard-wiechert}
\frac{e}{c}\frac{1}{R}\frac{\hat{n}\times((\hat{n}-\hat{v})\times\hat{a})}{(1 - \hat{v}\cdot\hat{n})}.
\end{equation}
In cases of linear acceleration such as ours, $\hat{v}\times\hat{a}=0$, and so $\hat{n}\times(\hat{n}\times\hat{a})$ produces a $\sin\t$ term. Therefore for an instantaneous acceleration we retain only dipole EM radiation, and analogously only quadrupole GW radiation, relative to the instantaneous position at retarded time. When the acceleration is prolonged rather than instantaneous, the position of the radiation source changes during the emission process. Therefore the radiation signal is a superposition of lowest-multipole terms from different times and sources - and thus includes higher multipoles. It is these contributions that we do consider\footnote{Higher multipoles can also appear if the acceleration is non-linear, i.e if the direction of $a$ is different than the direction of $v$. Then the numerator aquires a new term$-\hat{n}\times((\hat{v})\times\hat{a})$, which is different from the lowest multipole term. Such terms are irrelevant for our jet models, where the particles accelerate along straight lines. The method \refeq{lienard-wiechert} is used to calculate the EM Larmor formula (\cite{Jackson}), including only the instantaneous dipole term, and the GW Quadrupole Formula(\cite{Maggiore}).} here. In general, the calculation of the radiation generated by a matter distribution $\rho(t',{\bf x'})$ and seen by an observer at $(t,{\bf x})$ involves an integral of the form
\begin{equation} \label{generalwaveformintegral}
\frac{dh}{dt}^{jet}\!\!\!\!\!\! (t,{\bf x})\!=\!\!\!\int\!\!dt'\!\!\!\IIInt\!\!d^3{\bf x'}\d (t-t'-|{\bf x-x'}|)\rho(t',{\bf x'})\D{}h^{pp}({\bf x};{\bf x'}),
\end{equation}
where $\D h^{pp}$ is given by \refeq{hpointparticledeltah} for ${\bf r}={\bf x-x'}$, and the delta function enables us to calculate using the retarded time $t'_{ret}({\bf x},{\bf x'})=t-|{\bf x-x'}|$:
\begin{equation} \label{generalwaveformintegralretarded}
\frac{dh}{dt}^{jet}\!\!\!\!\!\! (t,{\bf x})=\IIInt d^3{\bf x'}\rho(t'_{ret},{\bf x'})\D{}h^{pp}({\bf x};{\bf x'}).
\end{equation}
Inspired by the Fireball model \cite{Fireball,PiranShemi}, we expect the jet to accelerate from $\Gm_i=1$ to $\Gm_f\sim100$, approximately linearly\footnote{This is of course an approximation, and might break down in a realistic model, for example when the jet is propagating within a Collapsar or if the acceleration is magnetic rather than thermally driven.}:
\begin{equation} \label{linearacc}
\Gm \propto r(t)\sim t.
\end{equation}
We use the total acceleration time $T_f\sim30ms$. For simplicity we assume acceleration to be angularly uniform over the jet ($\t<\t_0$):
\begin{equation} \label{acceleration}
\Gm=1+a\cdot t \tab 0\leq t \leq T_f,\,\,\, a=\frac{\Gm_f-\Gm_i}{T_f-T_i}\sim\frac{\Gm_f}{T_f}.
\end{equation}
At every instant $t$, we assume that all the ejected mass/energy is at the same radius $r(t)$ from the center of the GRB, with the entire envelope-sector (\refig{jetdiagram})\footnote{or rather partial envelope, for a uniform jet. For a non-uniform jet, such as the structured jet we examine, the mass density varies over the angle from the jet axis, but the envelope-front still all shares the same $r(t)$} expanding radially at the same (angularly uniform) instantaneous velocity $\beta(t)$ and Lorentz factor $\Gm(t)$. This model approximates the Radiative Fireball model \cite{PiranShemi}, where $\Gm \propto r\sim t$. This allows us to simplify \refeq{generalwaveformintegral} by requiring\\
\begin{equation} \label{singleshell}
\rho(t',{\bf x'})=f(\t',\phi')\d\left(r'-\int dt'\bt(t')\right),
\end{equation}
using the jet-model functions $f$ as in \refeq{deltahjet}. Thus
\begin{equation} \label{waveformintegral}
\frac{dh}{dt}^{jet}\!\!\!\!\!\! (t,{\bf x})\!\!=\!\!\!\IInt\!\!d cos\t'd\phi' f(\t',\phi')r_{ret}^2(\t',\phi',\t,t)\D h^{pp}(\t',\phi',\t),
\end{equation}
where now the integration is performed over all the matter elements responsible for the signal arriving at ${\bf x}$ at time $t$, with $r_{ret}(\t',\phi',\t,t)$ describing the distance of each source element from the origin.\\
Rather than calculating \refeq{waveformintegral} by explicitly solving \refeq{singleshell} and subtituting $r_{ret}(\t',\phi',\t,t)$ in the integral, we have taken a different numerical route based on our earlier calculations. We divide the source in space over an angular grid ($5000\times100$) of point-sources, and followed each such source as it accelerates (over 50000 time steps). Each point-mass accelerates linearly, maintaining its angular position $(\t,\phi)$, but at each time step its $r$,$\bt$ and $\Gm$ increase ($\Gm(t)$ linearly by \refeq{acceleration}, $\bt(t)$ matches $\Gm(t)$, and $r(t)$ as a time integral over $\bt(t)$). For each such point-mass at each time step, we model its acceleration as the disapperance of an incoming slow particle, and the appearance of an outgoing faster one. We sum both signal contributions using \refeq{hpointparticle1}, the positive (outgoing) contribution and the negative (incoming) one. We record the net signal along with its time of arrival at the observer. The final observed signal at time $t$ is the sum of the contributions arriving at that instant.\\
Since $\Delta h \propto \Gm \propto t$ (\refeq{hpointparticledeltah} and \refeq{acceleration}), each consecutive contribution of the envelope to the signal increases with time (until the jet reaches $\Gm$). The GW signal is ``anti-beamed", outside an angle $\t\sim\Gm^{-1/2}$ away from the direction of acceleration, with $\Gm \propto t$, the signal emitted at $t$ is anti-beamed outside an angle $\t\sim t^{-1/2}$. In the forward directions, where $\gm$-ray radiation can be detected, we find that the GW signal contribution is larger at later times (when $\Gm^{-1/2}$ is small enough as not to encompass the observer), while the siganl contributions emitted earlier (when the observer's angle was $\lesssim\Gm^{-1/2}$) is undetectable. An observer in the forward ($\gm$-ray) cone would therefore miss the early signal, and only catch the later parts. The reverse-jet also contributes to the late signal (fig.(\ref{waveformsdeltah}c)), as a shallow increase arriving later (until $\sim2T_f$). However this signal is much weaker than the forward one ($\sim10^{-4}$ compared with the forward signal, due to the beaming factor ($1/(1-\bt cos\t)$).\\
We expect to see the signal from the same Equal Arrival Time Surfaces (EATS) studied for EM radiation (\cite{Rees,Waxman,Granot}). These are characterized by two time-contraction effects: radial time contraction (forward) and angular time contaction (sideways)\cite{Piran}. The emitting jet travels at almost the speed of light (in the observer's frame), so in the forward line of sight (FLOS) the jet and the signals it emits almost catch up with the signals emitted earlier. Thus the time delay between observing signals emerging at times $t_1,t_2$ is contracted to
\begin{equation} \label{frontshock}
\Delta t\sim\frac{1}{2\Gm^2}(t_2-t_1).
\end{equation}
Similarly (as for $\gm$-ray waveforms \cite{Piran}), the parts of the jet expanding at an angle off the FLOS are more distant from the observer relative to the parts of the jet along the FLOS. This introduces a time-delay that depends on the angle of the source (increasing away from the FLOS), and increases linearly with t, as ct is approximately the instantaneous radius of the envelope:
\begin{equation} \label{sideshock}
\Delta t\sim(t_2-t_1)(1-\cos\t)\sim\frac{\t^2}{2}(t_2-t_1)
\end{equation}
in the arrival times between the signals from two sources emitting the signal simultaneously, on and of the FLOS. These two effects can be geometrically expressed together as
\begin{equation} \label{bothshocks}
\Delta t\sim\frac{\t^2+\Gm^{-2}}{2}(t_2-t_1).
\end{equation}
However, unlike the case of EM radiation, contributions to the GW signal from the FLOS or close to it are undetectable due to ``anti-beaming''. Thus $\t\gtrsim\Gm^{-1/2}$ for any contribution detectable by the observer, and the contraction factor is always $O(\Gm^{-1})$ rather than $O(\Gm^{-2})$.\\

\begin{figure}[h]
\includegraphics[width=86mm]{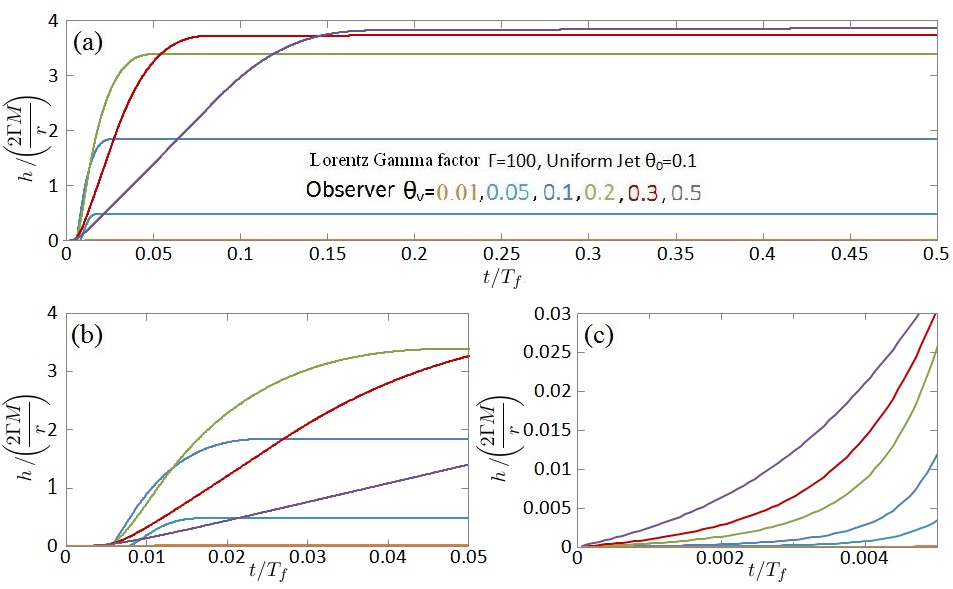}
\caption{h(t) for different observeration angles $\t_v$ (color-coded);$\Gm=100$, $\t_0=0.1$. Time is measured in dimensionless units $t/T_f$, and $h$ is scaled to $h/(\frac{2\Gm M}{r})$, where $\frac{2\Gm M}{r}\sim10^{-25}$ (compare \refeq{hpointparticledeltah},(\ref{htypicalamp}),(\ref{hpeak_pp})). (a) The entire waveform reaches plateau extending to $2T_f\sim60ms$. (b) A close-up of the high-rising signal from the forward jet, around $\Gm^{-1}T_f$. (c) The very early signal, visible sooner away from the line of acceleration.}
\label{waveformstotalh}
\end{figure}

\begin{figure}[h]
\includegraphics[width=86mm]{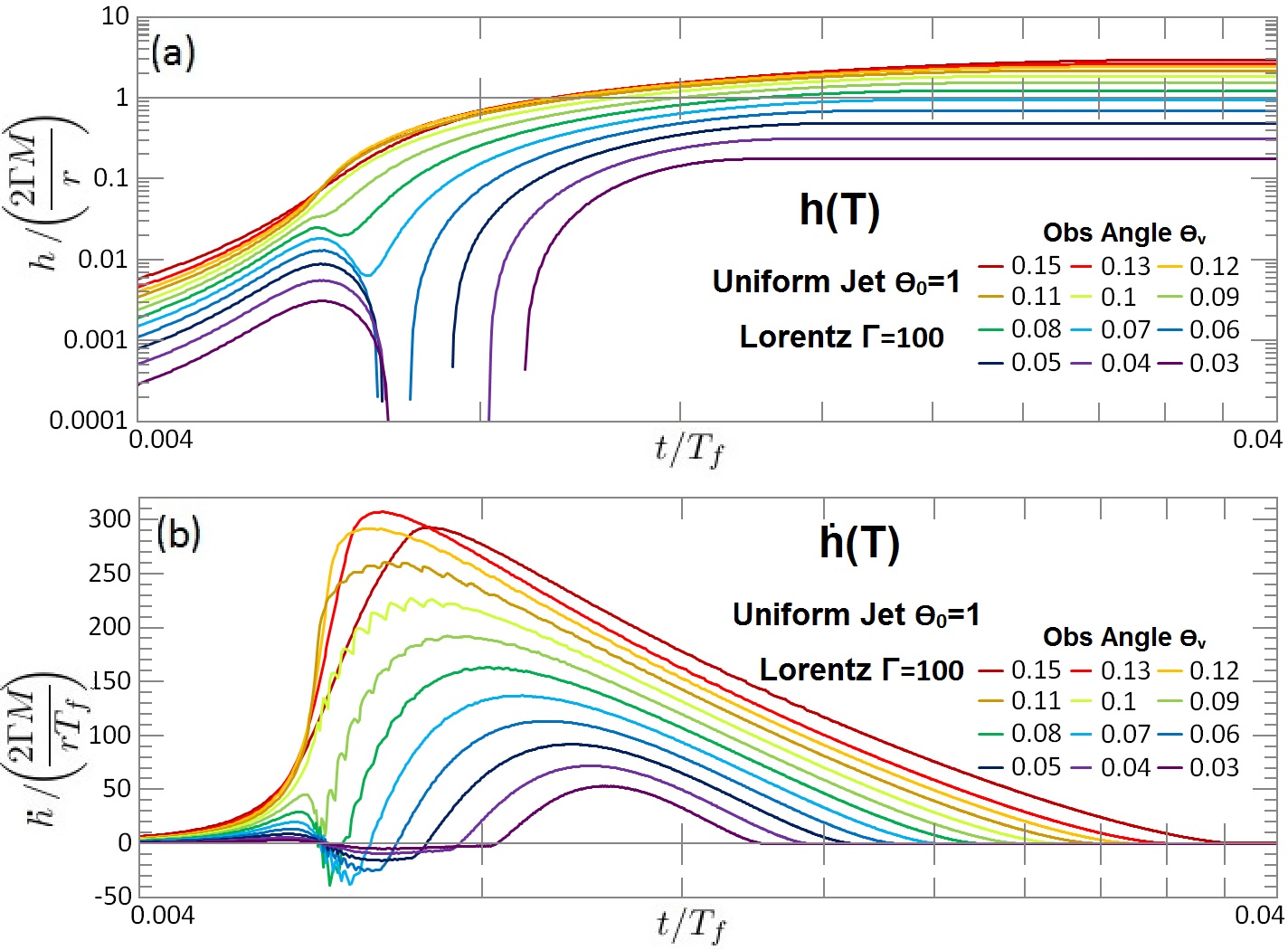}
\caption{GW signal from uniform jets of $\t_0=0.1$, $\Gm=100$, for different observation angles $\t_v$, around 0.1. (a) shows $h(t)$ and (b) shows $\dot{h}(t)$. For $\t_v<\t_0$, shortly after the signal buildup begins $\dot{h}$ dips, reaching negative values. For sufficiently low $\t_v$, $h$ itself becomes negative ($\t_0=0.06,0.05,0.04,0.03$). Later, the signal rises again. For $\t_v\sim\t_0$ phase-variations causes wiggles to appear around the peak (seen for $\t_0=0.08,0.09,0.1,0.11$)}
\label{waveformbumps}
\end{figure}

\begin{figure}[h]
\includegraphics[width=86mm]{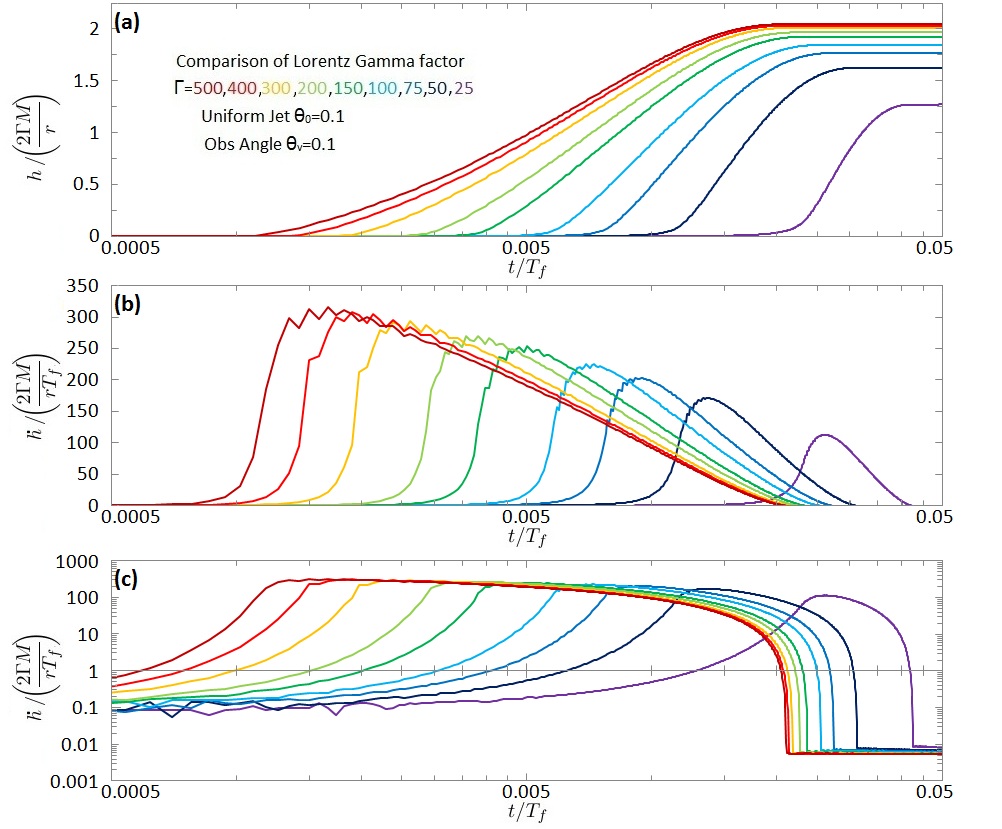}
\caption{h(t) (a) and $\dot{h}$ ((b) - linear scale, (c) - log scale) for different Lorentz $\Gm$ values (500,400,300,200,150,100,75,50,25) ($\t_0=0.1$, $\t_v=0.1$). The signal is stronger, and peaks sooner, the higher the Lorentz factor. $h$ is scaled to $h/(\frac{\Gm M}{r})$, and $\dot{h}$ to $\dot{h}/(\frac{\Gm M}{rT_f})$}
\label{waveformsgamma}
\end{figure}

\begin{figure}[h]
\includegraphics[width=60mm]{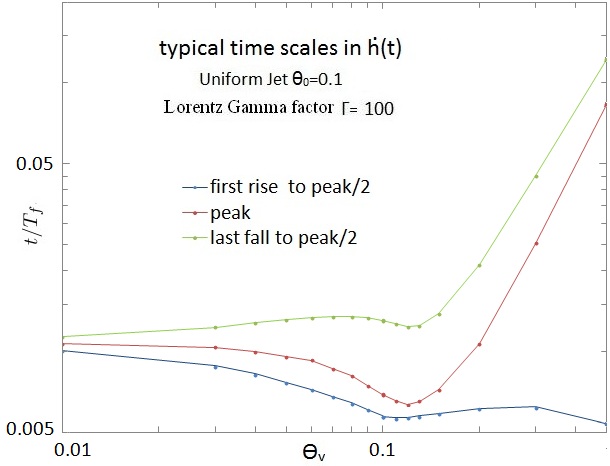}
\caption{A comparison of timescales for $\frac{dh}{dt}$ from a uniform jet of $\t_0=0.1,\Gm=100$ as a function of the observation angle $\t_v$. Shown are the times of the peak of $\frac{dh}{dt}$, as well as the times when it reaches $1/2$ of its peak value, both on the rising and on the falling edge.}
\label{timescales}
\end{figure}

\subsection{The Waveform}				
\begin{figure}[h]
\includegraphics[width=86mm]{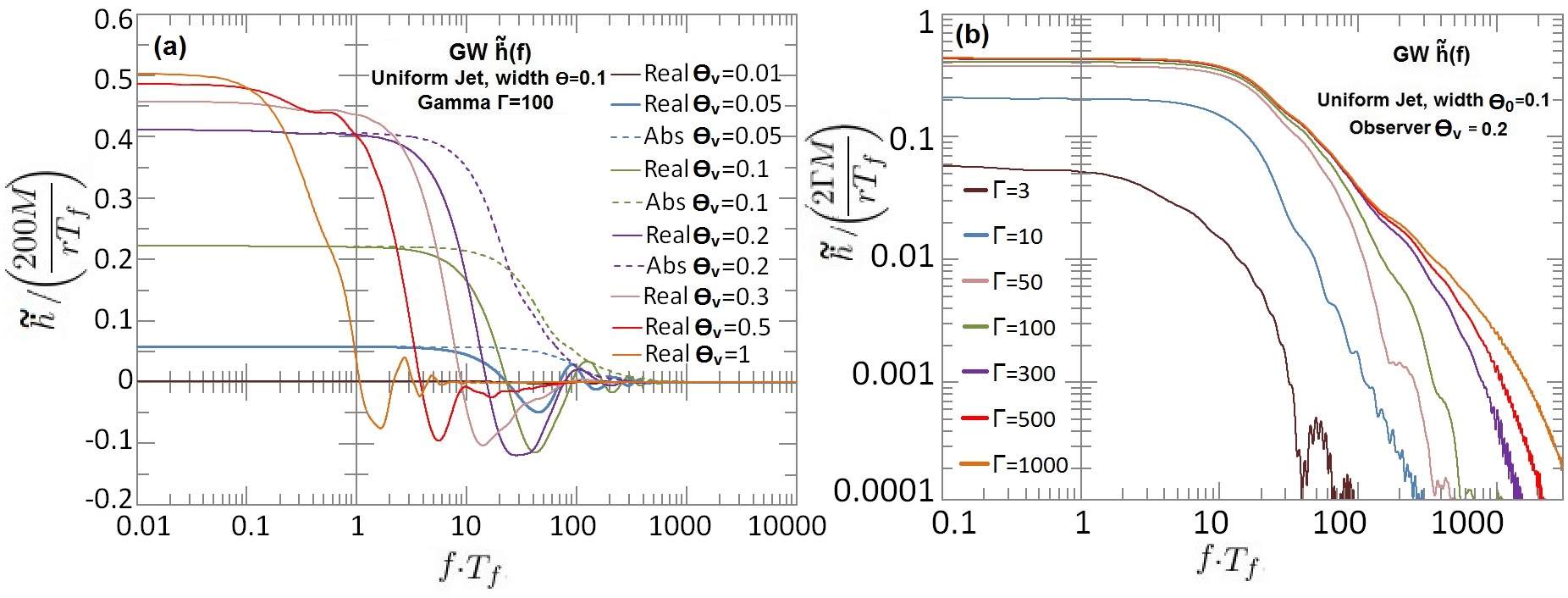}
\caption{Fourier analysis of the waveforms produced by (a) uniform jets ($\Gm=100$,$\t_0=0.1$) viewed from different angles $\t_v$ and (b) uniform jets ($\t_0=0.1$, viewed from $\t_v=0.2$) for different $\Gm$ factors. Time is plotted in dimensionless units $t/T_f$, and the Fourier spectrum is scaled to $\tilde{\dot{h}}/(\frac{2\Gm M}{rT_f})$. All patterns match sinc functions.}
\label{Fourieruni}
\end{figure}

\begin{figure}[h]
\includegraphics[width=60mm]{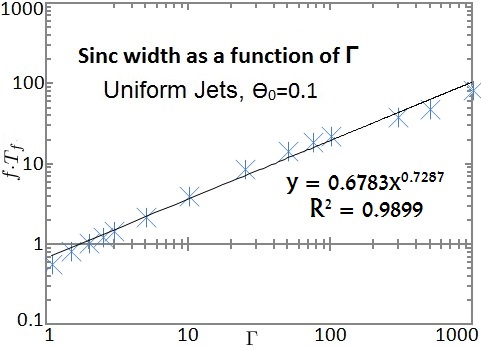}
\caption{$F$, the width (first zero) of the Fourier transform of $\dot{h}$, as a function of $\Gm$, for $\t_v=\t_0=0.1$.}
\label{FourierwidthG2}
\end{figure}

\begin{figure}[h]
\includegraphics[width=86mm]{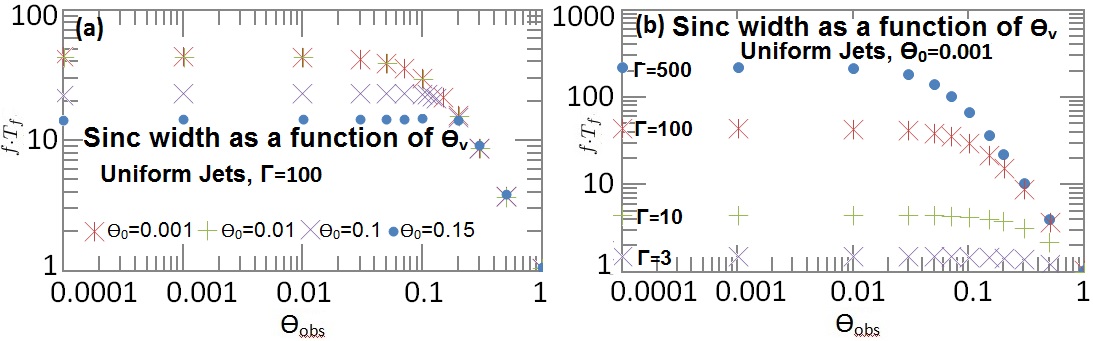}
\caption{$F$, the width (first zero) of the Fourier transform of $\dot{h}$ as a function of $\t_v$, for (a) $\Gm=100$ and different jet widths; and (b) for different $\Gm$'s with width$\t_0=0.001$.}
\label{Fourierwidthboth}
\end{figure}

\begin{figure}[h]
\includegraphics[width=86mm]{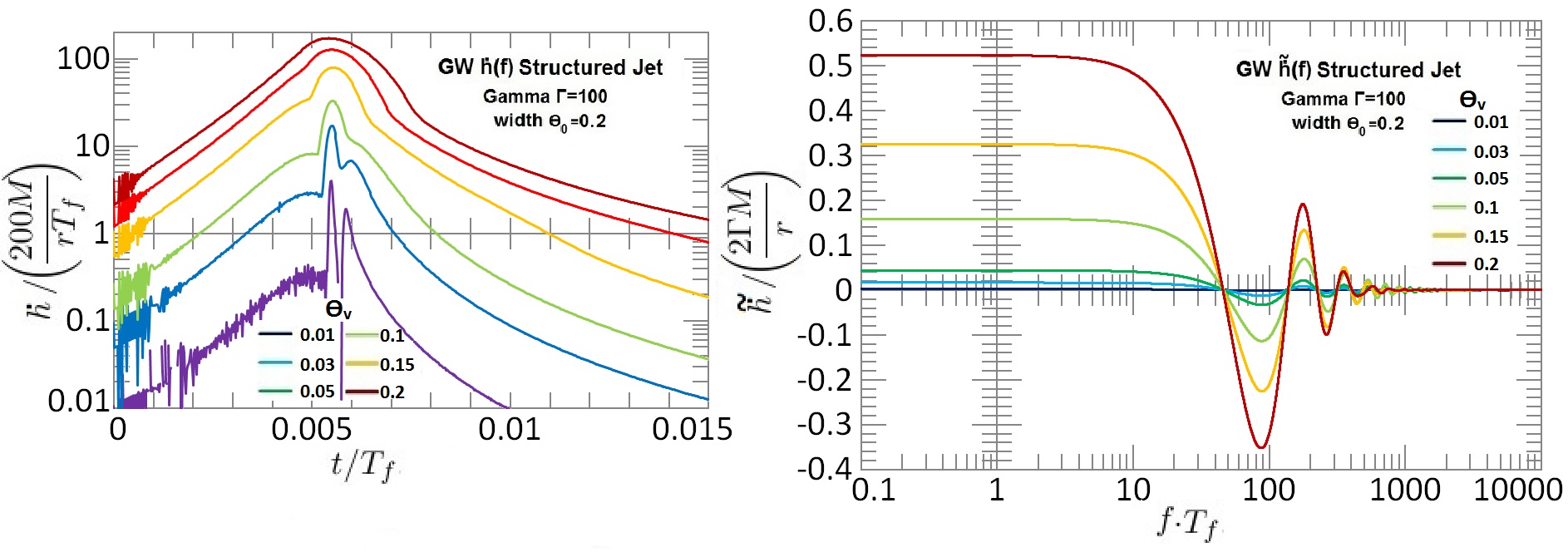}
\caption{(a) $\dot{h}$ as a function of time for a structured jet ($\Gm=100$, $\t_0=0.2$) at different observeration angles $\t_v$. We see a peak pulse similar to a narrow uniform jet, and at very low angles 
$\t_v\sim\t_{core}\sim\Gm^{-1}$ we find the pulse is followed by a negative dip. (b) Fourier analysis of the same waveforms produced by structured jets ($\Gm=100$,$\t_0=0.2$) viewed from different angles. The patterns match sinc functions.} 
\label{waveformsFouriertructured}
\end{figure}

We have run simulations of a uniform jet with a constant, uniform linear acceleration, over a wide parameter space of
$\Gm,\t_0,\t_v$. We have used different numbers of integrations steps to confirm the numerical validity of the results. The resulting waveforms demonstrate the expected effects. $T_f$ factors out linearly, and the waveform results are identical (up to scaling) for bursts of different durations which reach the same final $\Gm$; prolonging or shortening the acceleration is equivalent to stretching the waveforms in time - or shifting them in Fourier space. We describe here the results for a characteristic $T_f=30ms$. Figures (\ref{waveformstotalh},\ref{waveformsdeltah},\ref{waveformbumps}) show the waveform from different observation angles for a uniform jet of parameters $\Gm=100$,$\t_0=0.1$.\\
Fig.(\ref{waveformstotalh}a) depicts the double-feature of the aggregated signal, from the forward and reverse jets. First, the contracted forward jet's signal (fig.(\ref{waveformsgamma}c)) rises rapidly over a time scale $t_{rise}\sim\Gm^{-1}T_f$, matching our anticipation (\refeq{bothshocks} and on). This is most clearly visible for small observation angles ($\t_v\lesssim\t_0$), where the jet width sets the angular time-scale; for larger angles, the peak and total duration timescales are set by $\t_v>\t_0$ (\refig{timescales}). It is followed by the long ($\sim2T_f\sim60ms$) and weaker (by $\sim10^{-4}$) backward-jet-signal (fig.(\ref{waveformsdeltah}c)). Fig.(\ref{waveformsdeltah}b) demonstrates clearly the anti-beaming effect and its dependance on time (via $\Gm(t)$). The onset of the signal for the different observers is determined by the observation angle $\t_v$: The most off-axis observers observe the signal first. As $\Gm$ increases with $t$, relativistic beaming wins over, and observers closer to the forward axis detect the signal.\\
Fig.(\ref{waveformsdeltah}b) depicts the angular spreading of the forward signal. We find that for our characteristic $T_f=30ms$ burst, the strong signal from the forward jet lasts for $\sim0.2-0.3ms$. This angular spreading timescale ($\sim0.5(\t_0)^2T_f$) is longer than the timescale of the forward time-contraction ($\sim0.5\Gm^{-2}T_f$), and thus it masks the latter, as explained following \refeq{bothshocks}. This result of GW anti-beaming should be contrasted with the beaming of the EM signal, for which the forward signal is significant, and thus the radial time contraction sets the timescale. Fig. (\ref{waveformsgamma}) makes more apparent (for the same uniform jet openning angle and the same observer angle) the $\Gm$-dependence of the effects: a larger $\Gm$ produces a stronger signal, which is also more strongly beamed for the forward observer (and thus reaches him sooner), and more time-contracted.\\
A different acceleration mechanism - in particular, acceleration driven by Poynting flux rather than a thermal fireball - would change the relation between $\Gm,r$ and $t$, and thus change the angular beaming and time contraction characters of the signal. These should all produce different waveforms, with implications on detectability. Thus from the GW signal we can learn about the acceleration of the jet and about their angular distribution. This could distinguish between uniform, structured and shotgun\cite{Heinz} jets (all have different angular distributions) and possibly (after further study) between Poynting flux acceleration and thermal fireball acceleration.\\
We notice an interesting feature of the wavforms observed from $\t_v<\t_0$ (for example 0.05 in \refig{waveformsdeltah}): shortly after the signal $\dot{h}$ begins to rise, it drops sharply, reaching negative values, and only then it rises again. We observe this feature in a range of waveforms for various values of $\t_v$,$\t_0$ (\refig{waveformbumps}). It is produced by the region of the jet surrounding the direction to the observer. In this region, the angle $\phi$ around the FLOS produces phase factors $e^{2i\phi}$ (\refeq{hpointparticledeltah}) which include negative contributions to $\dot{h}_+$. Since this negative-dip arrives from directions close to the FLOS, it precedes the main signal. For sufficiently small observing angles, we find time periods where even the total signal $h_+$ becomes negative (\refig{waveformbumps}). A similar effect of phase-variation from angles close to the FLOS causes, for $\t_v \sim \t_0$, the appearance of wiggles around the peak, as verified consistently for many waveforms and numerical simulations (\refig{waveformbumps})\\
By Fourier-transforming the waveforms from uniform jets (\refig{Fourieruni}), we find that the most prominent feature is a $sinc$ function, which matches our expectation of transforming a pulse in the time domain: the Fourier components are approximately constant starting at very low frequencies and then they drop to zero. The width of the $sinc$, which we measure as the first zero of the function, corresponds to the width of the pulse in the time domain, and we find it at a frequency range $f_0\sim600Hz$ (for $T_f=30ms$\footnote{As mentioned, in our model $T_f$ in fact factors out linearly, and so all frequency results scale linearly with $1/T_f$}, and depending on the viewing angle, jet width and Lorentz factor). For a uniform jet of $\t_0=\t_v=0.1$ we find (\refig{FourierwidthG2})
\begin{equation}
\label{typicalfreq}
f_0\sim\frac{0.67}{T_f}\Gm^{0.73}.
\end{equation}
We find that at larger viewing angles $f_0$ is independant of $\Gm$ and $\t_0$, while for lower values of $\t_v$ it does not depent on $\t_v$ (\refig{Fourierwidthboth}). For $\t_v<\t_0$, we see a second (higher) range of frequencies for which the spectrum is non-negligible, matching the negative-dip in the waveforms for these angles.\\
Fig.(\ref{waveformsFouriertructured}a) shows the waveforms for a structured jet ($\Gm=100$,$\t_0=0.2$) from several observation angles, displaying similar features to a narrow uniform jet. It's Fourier spectrum (fig.(\ref{waveformsFouriertructured}b)) matches a sinc-pulse, regardless of the observation angle, indicating most of the signal arises from the very narrow uniform core. We also see that for $\t_v\sim\t_{core}\sim\Gm^{-1}$, the signal includes a negative dip which follows the main signal rise, rather than precede it as in the uniform signature.

\begin{table}[h!]
\centering \caption{SNR for GW from GRB ($\Gm M=10^{51}erg$, at $r=500Mpc$) relative to expected noise in aLIGO \& DECIGO. For Advanced LIGO we use the two baselines ZERO-DET-high-P and ZERO-DET-low-P\cite{LIGO}. For DECIGO we use three baselines: the expected sensitivity for a single cluster\cite{DECIGO2}, a fiducial DECIGO-like detector of sensitivity $\sim10^{-25}/\sqrt{Hz}$ at the DECIGO band, and the ultimate DECIGO sensitivity\cite{DECIGOoriginal}.}
\label{table:SNR}
\begin{center}
\begin{tabular}{ccc|ccccc}
\hline\hline
	\multicolumn{2}{c}{Jet}&$\theta_{obs}$ &\multicolumn{2}{c}{\bf Adv LIGO}&\multicolumn{3}{c}{\bf DECIGO}\\
	&&&\multicolumn{2}{c}{ZERO-DET}\\
Model&$\Gamma$&&high-P&low-P&1-cluster&fiducial&ultimate\\
\hline
	STR &100	&	0.05	&	8E-03	&	4E-03	&	{\blue 1E-01} &	5E+00	&	{\blue 3E+02}\\
	STR &100	&	0.1		&	3E-02	&	2E-02	&	{\blue 4E-01} &	2E+01	&	{\blue 1E+03}\\
	STR &100	&	0.15	&	6E-02	&	3E-02	&	{\blue 8E-01} &	4E+01	&	{\blue 2E+03}\\
	STR &100	&	0.2		&	9E-02	&	5E-02	&	{\blue 1E+00} &	6E+01	&	{\blue 3E+03}\\
	STR &100	&	0.3		&	2E-01	&	9E-02	&	{\blue 2E+00} &	1E+02	&	{\blue 5E+03}\\ \hline
	UNI &100	&	0.01	&	4E-04	&	2E-04	&	{\blue 6E-03} &	3E-01	&	{\blue 2E+01}\\
	UNI &100	&	0.05	&	9E-03	&	5E-03	&	{\blue 1E-01} &	6E+00	&	{\blue 3E+02}\\
	UNI &100	&	0.1		&	4E-02	&	2E-02	&	{\blue 4E-01} &	2E+01	&	{\blue 1E+03}\\
	UNI &100	&	0.2		&	6E-02	&	3E-02	&	{\blue 8E-01} &	4E+01	&	{\blue 2E+03}\\
	UNI &100	&	0.3		&	5E-02	&	3E-02	&	{\blue 1E+00} &	5E+01	&	{\blue 3E+03}\\		
	UNI &100	&	0.5		&	3E-02	&	2E-02	&	{\blue 1E+00} &	5E+01	&	{\blue 3E+03}\\
	UNI &100	&	1		&	7E-03	&	7E-03	&	{\blue 1E+00} &	5E+01	&	{\blue 3E+03}\\ \hline
	UNI &1000	&	0.05	&	1E-02	&	6E-03	&	{\blue 1E-01} &	7E+00	&	{\blue 4E+02}\\
	UNI &1000	&	0.1		&	4E-02	&	2E-02	&	{\blue 6E-01} &	3E+01	&	{\blue 2E+03}\\
	UNI &1000	&	0.2		&	7E-02	&	4E-02	&	{\blue 1E+00} &	5E+01	&	{\blue 3E+03}\\ \hline
	UNI &500	&	0.05	&	1E-02	&	6E-03	&	{\blue 1E-01} &	7E+00	&	{\blue 4E+02}\\
	UNI &500	&	0.1		&	4E-02	&	2E-02	&	{\blue 1E+00} &	7E+01	&	{\blue 4E+03}\\
	UNI &500	&	0.2		&	7E-02	&	4E-02	&	{\blue 1E+00} &	5E+01	&	{\blue 3E+03}\\ \hline
	UNI &300	&	0.05	&	1E-02	&	6E-03	&	{\blue 1E-01} &	7E+00	&	{\blue 4E+02}\\
	UNI &300	&	0.1		&	4E-02	&	2E-02	&	{\blue 6E-01} &	3E+01	&	{\blue 2E+03}\\
	UNI &300	&	0.2		&	6E-02	&	4E-02	&	{\blue 1E+00} &	5E+01	&	{\blue 3E+03}\\ \hline
	UNI &50		&	0.05	&	8E-03	&	5E-03	&	{\blue 1E-01} &	6E+00	&	{\blue 3E+02}\\
	UNI &50		&	0.1		&	3E-02	&	2E-02	&	{\blue 4E-01} &	2E+01	&	{\blue 1E+03}\\
	UNI &50		&	0.2		&	5E-02	&	3E-02	&	{\blue 8E-01} &	4E+01	&	{\blue 2E+03}\\
\hline\hline
\end{tabular}
\end{center}
\end{table}

\sect {Conclusions}							%
We studied the gravitational waves radiated from the accelerating GRB jets, examining both uniform jets and structured jets. We found that the quadrupole nature of GW causes anti-beaming away from the jet's axis, and thus the radiation peaks approximately at an angle $\Gm^{-1/2}$ off-axis. For a uniform jet, the signal is maximal just-outside the jet; for a structured jet, outside the core, but within the jet. For $\t_0>\Gm^{-1}$, the structured jet's GW signal is always more focused than its uniform counterpart, and as $\t_0$ increases, it remains more focused than the uniform jet signal (\refig{jetantena}, (\ref{both})). As $\gamma$-rays from the jet are emitted within the jet width $\t_0$, the GW peak is generally not within the GRB detectability cone; however the GW amplitude decreases only slightly away from its peak, and so a considerably high amplitude can be seen within the EM cone (\refig{jointdetection}).\\
The waveform and the spectrum of the GW signal depend on the acceleration of the jet. We introduced an acceleration model of uniform linear acceleration, that is based on the Fireball model for thermal acceleration (\refeq{linearacc}). It might be interesting to explore in the future other acceleration mechanisms and in particular the acceleration of a Poynting flux dominated jet. As the relations between $\Gm$, $r$ and $t$ determine the beaming and time contractions, the different GW signals could teach us about the acceleration of the jetand could help distinguish between Poynting flux acceleration and thermal fireball acceleration.\\
For a source at distance $r$, of total energy $M$ and a bulk Lorentz factor $\Gm$, that accelerates over time $T_f$, the signal's amplitude scales linearly with $\frac{\Gm M}{r}$, and the waveform stretches over time linearly with $T_f$. The signal increases, first sharply over a timescale $t_{rise}\sim0.5\Gm^{-1}T_f$. this is followed by a long slow increase until $\sim2T_f$ (figs (\ref{waveformstotalh},\ref{waveformsdeltah},\ref{waveformbumps},\ref{waveformsgamma})). The fast rise time represents a time contraction of $\Gm^{-1}$ stemming from the angular quadrupolar anti-beaming of the GW, which stands in contrast with the $\Gm^{-2}$ time contraction factor of the EM radiation from the GRB. Another effect, we found, is the appearance of wiggles in the signal, preceding its peak (figs (\ref{waveformbumps},\ref{waveformsgamma})) , at observation angles within the jet cone, due to polarization. This implies that for a uniform jet the signal is much clearer and monotonous outside the jet's cone than the signal of a structured jet; however it also means that for any uniform jet detectable as a GRB (i.e., the observer is within the jet) we expect to see such wiggles in the GW signal, and they are more pronounced the closer the observe is to the jet axis. The waveform of a structured jet displays a similar dip after the signal's peak. Thus the waveforms, and particularly the wiggles, offer insight into the internal structure of the jet, in particular to the angular structure and the acceleration.\\
We turn now to analyze the detectability and exptected detection event rate using the frequency bands and sensitivities of the aLIGO and DECIGO detectors. Advanced LIGO \cite{LIGO} is planned to detect strain sensitivities of $\sim10^{-24}-10^{-23}/\sqrt{Hz}$ in the range $30-300Hz$, while DECIGO \cite{DECIGO,DECIGO2} should reach the same sensitivity in the $0.1-10Hz$ range. The characteristic frequencies of the GRB jet's GW's scale with $1/T_f$ (\refeq{typicalfreq}). We have calculated the Fourier signatures of different waveforms, and found their main feature is a $sinc$ funtion (figs \ref{Fourieruni}, \ref{FourierwidthG2}, \ref{Fourierwidthboth}, \ref{waveformsFouriertructured}). For a typical uniform jet of width $\t_0=0.1$, $\Gm=100$ and acceleration time $T_f=30ms$ viewed at $\t_v=0.1$, we find the width of the $sinc$ is $f_0\sim20Hz\Gm^{0.73}\sim600Hz$. Following Flanagan et el \cite{Flanagan,Helstrom,Wainstein,Blair,Saulson}, we have computed the Signal-to-Noise Ratio (SNR) for our waveforms using the expected sensitivity curves of aLIGO (with two operation-mode baselines: ZERO DET high P \& ZERO DET low P\cite{LIGO}) and DECIGO ({\blue with three baselines: the expected single cluster sensitivity, the ultimate sensitivity from the original DECIGO paper\cite{DECIGOoriginal}, and an intermediate fiducial sensitivity $10^{-25}/\sqrt{Hz}$ in the DECIGO band\cite{DECIGO2})}, using
\begin{equation} \label{SNReq}
\rho=\left(\frac{S}{N}\right)_{optimal filter}=2\sqrt{\int\frac{|\tilde{h}(f)|^2}{S_h (f)}df}.
\end{equation}
We used a typical parameter of total energy $M=10^{51}erg$, and considered a fiducial source at $r=500Mpc$, for comparison with the gravitaional radiation estimaes of \cite{StatisticsPiran}.\\
{\blue
The results are shown in table \ref{table:SNR}. Using DECIGO's ultimate sensitivity, we see that the SNRs are large enough to allow detection to several orders of magnitude farther than our fiducial distance, even to high cosmlogical z (We note the SNR scales linearly with $M$ and inversely with $r$ when z is small, and similar even for high z). With our less optimisitc fiducial sensitivity, we find SNRs in the range 5-100 depending on the jet structures (table \ref{table:SNR}). For a canonical uniform jet of width $0.1$, the SNR (20) allows detection up to $\sim4$ time farther ($~2Gpc$). Using the local GRB rate estimates of $\sim1.3Gpc^{-3}yr^{-1}$ for long GRB's \cite{StatisticsWanderman}, we expect a monthly joint detection rate. The rate of short GRB's is higher, $\sim8Gpc^{-3}yr^{-1}$ \cite{StatisticsCoward}, but their typical energy is $~10$ times smaller, reducing their GW's detection range by $~sim10$; taken together, we expect a detection of GW from a short GRB about once a decade. We also expect to see GW signals from jets without seeing the corresponding GRB itself, due to their different beaming properties: the GW's amplitude is considerable outside a $\sim\Gm^{-1/2}$ cone, $\sim50$ times wider than the $\gm$-ray angle\footnote{The ratio between a GW solid angle of almost $4\pi$ to the GRB solid angle of $\sim2\pi\t_0^2$, and assuming a narrow GRB opening angle of $\t_0\sim0.1-0.2$ \cite{StatisticsWanderman}.}. For the signals coming from jets pointing towards us at angles that are of order $\Gm^{-1/2}$ we expect significant orphan afterglow \cite{StatisticsPiran} signals to accompany the GW signals.\\
A GRB jet's GW is also expected to be preceded by GW from its progenitor, and the coincidence of the two signals (even on different frequency scales and using different detectors) could improve the chances of finding them.\\
Using aLIGO or a single DECIGO cluster and our typical parameters, the SNR for a uniform jet is expected to be $\sim1-5\cdot10^{-2}$, while for a structured jet it can reach $10^{-1}$; both are undetectable. A GW signal can be detected together with a GRB if it is $\sim10$ times stronger/closer, thus $\sim1000$ times less frequent ($\sim10^{-3}yr^{-1}$). This rate is prohibitively low even considering isotropic GW signals (whose GRB and/or orphan afterglow counterparts are not visible).}\\
Other astrophysical phenomena might produce similar GW signals. In particular, Micro Quasars ejecting mass and energy at relativistic velocities are less powerful ($\Gm M\sim10^{44}erg$), but they are much closer to us ($r\sim5Kpc$). Using \refeq{htypicalamp}, we estimate their peak to be $h\sim10^{-27}$, two orders of magnitude less than the typical GRB signals; we expect their typical frequencies to be $f\lesssim35Hz$ (for a similar $T_f$ but much smaller $\Gm\sim2$, compare \refeq{typicalfreq},\refig{FourierwidthG2}). Active Galactic Nuclei can also display acceleration (magnetic) and emit much more energy, over longer time scales and lower frequencies. The same techniques, once adapted to continous rather than impulsive sources, could be applied to analyze the GW output of such sources.\\

\begin{acknowledgments}
The research was supported by an ERC Advanced grant. We thank Amos Ori for helpful discussions and comments.
\end{acknowledgments}

\appendix

\section {Additional Resources}
All of the code used is available at http://www.phys.huji.ac.il/$\sim$ofek/GWGRB/


\end{document}